\documentclass[twocolumn]{aastex631}
\usepackage{amsmath}
\usepackage{amssymb}
\usepackage{apjfonts}

\newlength{\apjcolwidth}
\shortauthors{Bauer et al.}

\begin{document}

\title{Masses of White Dwarf Binary Companions to Type~Ia Supernovae Measured from Runaway Velocities}

\correspondingauthor{Evan B. Bauer}
\email{evan.bauer@cfa.harvard.edu}

\author[0000-0002-4791-6724]{Evan~B.~Bauer}
\affiliation{Center for Astrophysics $\vert$ Harvard \& Smithsonian, 60 Garden St, Cambridge, MA 02138, USA}

\author[0000-0002-0572-8012]{Vedant~Chandra}
\affiliation{Center for Astrophysics $\vert$ Harvard \& Smithsonian, 60 Garden St, Cambridge, MA 02138, USA}

\author[0000-0002-9632-6106]{Ken~J.~Shen}
\affiliation{Department of Astronomy and Theoretical Astrophysics Center, University of California, Berkeley, CA 94720, USA}

\author[0000-0001-5941-2286]{J.~J.~Hermes}
\affiliation{Department of Astronomy \& Institute for Astrophysical Research, Boston University, 725 Commonwealth Ave., Boston, MA 02215, USA}

\begin{abstract}
  The recently proposed
  ``dynamically driven double-degenerate double-detonation'' (D$^6$)
  scenario posits that Type~Ia supernovae (SNe) may occur during
  dynamically unstable mass transfer between two white dwarfs (WDs) in a binary.
  This scenario predicts that the donor WD may then survive
  the explosion and be released as a hypervelocity runaway, opening up
  the exciting possibility of identifying remnant stars from D$^6$ SNe
  and using them to study the physics of detonations that produce Type Ia SNe. Three
  candidate D$^6$ runaway objects have been identified in {\it Gaia}
  data. 
  The observable runaway velocity of these remnant objects 
  represents their orbital speed at the time of SN detonation.
  The orbital dynamics and Roche lobe geometry required in the
  D$^6$ scenario place specific constraints on the radius and mass of
  the donor WD that becomes the hypervelocity runaway. In this letter,
  we calculate the radii required for D$^6$ donor WDs
  as a function of the runaway velocity.
  Using mass-radius relations for WDs, we
  then constrain the masses of the donor stars as well.
  With measured velocities for each of the three D$^6$ candidate
  objects based on {\it Gaia} EDR3, this work provides a new probe of
  the masses and mass ratios in WD binary systems that produce
  SN detonations and hypervelocity runaways.
\end{abstract}


\section{Introduction}

The ``dynamically driven double-degenerate double-detonation'' (D$^6$)
scenario for producing Type~Ia supernovae describes a double white
dwarf (WD+WD) binary system in which dynamically unstable mass transfer
ignites a supernova (SN) detonation that fully disrupts the accreting WD,
possibly releasing the donor WD from the binary as a hypervelocity runaway star
at approximately its former orbital velocity \citep{Shen2018a,Shen2018}. Simulations have identified this scenario
as a promising mechanism for igniting detonations in carbon-oxygen (C/O) WDs below
the Chandrasekhar mass \citep{Guillochon2010,Dan2011,Dan2012,Raskin2012,Pakmor2012,Pakmor2013,Shen2014,Shen2018a,Tanikawa2018,Tanikawa2019,Gronow2020,Gronow2021}.
Additionally, the thin He envelope that sustains the initial He detonation 
can lead to an explosion that produces the spectral features of
normal Type~Ia SNe \citep{Polin2019,Townsley2019,Boos2021,Shen2021a,Shen2021}.

Using astrometry from {\it Gaia} DR2, \cite{Shen2018} identified
three candidate D$^6$ remnant objects, which we will refer to as
D6-1, D6-2 (also known as LP~398-9), and D6-3.
These hypervelocity stellar remnants all have Galactocentric
velocities in excess of $1000\,\rm km\,s^{-1}$.
The longer baseline of {\it Gaia} EDR3 provides data confirming the
status of these three objects as hypervelocity runaway stars with
greater precision. The velocities and trajectories of these three
objects are difficult to explain with any other mechanism than
liberation from a WD+WD binary system that is compact enough ($P_{\rm orb}<10$\,min) for
the orbital velocities to exceed $1000\,\rm km\,s^{-1}$.
It remains an open question what fraction of Type Ia SNe
(or perhaps even peculiar thermonuclear SNe) the D$^6$ channel may
contribute to (e.g.\ see recent review by \citealt{Soker2019}),
but the extreme velocities of the observed D$^6$ runaways clearly point
to at least some fraction of thermonuclear SNe occurring in
very compact WD+WD binaries.
Our focus in this letter is therefore to characterize the constraints
on the progenitor WD+WD binaries that likely produced the three runaway
objects observed by \cite{Shen2018}.

The D$^6$ scenario specifically predicts an orbital configuration in
which the donor WD fills its Roche lobe at the time of SN detonation
when it is liberated, so the orbital dynamics and geometry of such a
system place a strong constraint on the donor WD radius.
Since the Roche-filling donor must be a WD, the WD
mass-radius relation then constrains the possible mass of the
donor WD based on the measured velocities of the D$^6$ objects.

In this letter, we present updated velocity distributions for the
three candidate D$^6$ donor remnants based on {\it Gaia} EDR3 in
Section~\ref{s.velocities}. In Sections~\ref{s.radius} and~\ref{s.mass},
we then describe how these velocities can be used along with the WD
mass-radius relation to constrain the masses of these objects as donor
WDs with the geometry required by the D$^6$ scenario.
Surprisingly, none of the resulting mass inferences point to the three
candidate remnant objects being descended from standard $\approx$0.6~$M_\odot$
WDs. Instead, the relatively low velocity of D6-2 requires a large radius for its
configuration as a Roche-filling donor star that implies it would have been
$\lesssim$0.2~$M_\odot$ as a WD donor unless it was significantly inflated by tides,
while the extreme velocities of D6-1 and D6-3 suggest that they were most
likely $\approx$0.8--1.1~$M_\odot$ as donor WDs in the D$^6$ scenario. 
In Section~\ref{s.rotation} we explore the implications of our calculations for the current measured rotation period of D6-2,
and we discuss some further implications of our mass inferences in
Section~\ref{s.discussion}.

\section{Velocities of the D$^6$ Stars}
\label{s.velocities}

\begin{figure}
  \centering
  \includegraphics[width=\apjcolwidth]{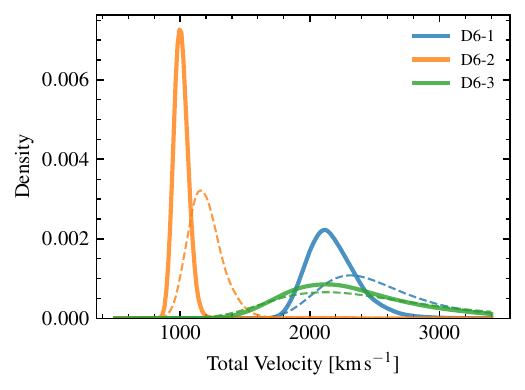}
  \caption{Posterior distributions of total heliocentric velocity
    for the D$^6$ stars, using astrometry from {\it Gaia} EDR3 (solid
    lines) and {\it Gaia} DR2 (dashed lines).}
  \label{fig:velocities}
\end{figure}

We compute updated velocities for the D$^6$ stars using
data from {\it Gaia} EDR3\footnote{The D$^6$ stars have \textit{Gaia} EDR3 Source ID 5805243926609660032 for D6-1, 1798008584396457088 for D6-2, and 2156908318076164224 for D6-3.} \citep{Gaia2021,Lindegren2021}. We sample
the parallax and proper motion distributions using the ensemble Markov
Chain Monte Carlo sampler \texttt{emcee} \citep{Foreman-Mackey2019},
assuming Gaussian uncertainties and taking into account the
covariances reported by {\it Gaia}. We apply an exponentially
decreasing space density prior on the distance, with a scale length
of 1350\,pc \citep[e.g.,][]{Astraatmadja2016}, and apply an empirical correction\footnote{\url{https://gitlab.com/icc-ub/public/gaiadr3_zeropoint}} to
the parallax zero-point \citep{Lindegren2021}. We utilize radial velocities from \cite{Shen2018} for D6-1 and D6-3,
and \cite{Chandra2021} for D6-2.
We display the posterior probability
distributions of the velocities in Figure \ref{fig:velocities},
along with the previous measurements from {\it Gaia} DR2 for
comparison. We estimate $1\sigma$ uncertainties using the 16th and 84th quantiles of the posterior samples. 
The final heliocentric velocities are $2160^{+200}_{-160}\, \rm km\,s^{-1}$ 
for D6-1, $1010^{+60}_{-50}\, \rm km\,s^{-1}$ for
D6-2, and $2270^{+600}_{-400}\, \rm km\,s^{-1}$ for D6-3.

In the D$^6$ scenario, the measured velocity of the WD
donor remnant is produced almost entirely by its orbital velocity $v_{\rm orb}$ at
the time of companion explosion, which is $\gtrsim 1000\, \rm
km\,s^{-1}$ in WD+WD binaries with a high-mass primary.
The donor WD is also expected to receive a kick from the ejecta with a 
magnitude of a few hundred $\rm km\,s^{-1}$ perpendicular to the
direction of orbital velocity, but this can only change the total
velocity by a few percent when added in quadrature with
the orbital velocity \citep{Wheeler1975,Taam1984,Marietta2000,Hirai2018,Bauer2019}.

The observed D$^6$ remnants lie within a few kpc of the Sun \citep{Shen2018},
and we expect that they were most likely ejected from progenitor binaries
near the Sun with roughly co-moving Galactic orbits.
Therefore, the measured velocities in the heliocentric frame are likely
the best representation of the orbital velocities at which they were ejected
from the progenitor binary. Accounting for alternative progenitor orbits
relative to local Galactic rotation could introduce an additional velocity
uncertainty of up to $\pm 200\,\rm km\,s^{-1}$ depending on the orientation
of observed runaway velocity.
For example, when compared to the velocities reported above the heliocentric frame,
converting to the Galactocentric frame using \texttt{astropy} \citep{AstropyCollaboration2018}
yields velocities that are $\approx 200\, \rm km\,s^{-1}$
slower for D6-1, $\approx 100\, \rm km\,s^{-1}$ faster for D6-2, and nearly unchanged
for D6-3. However, we adopt the heliocentric velocities for the remainder
of this work, as they are most likely to best represent the orbital
ejection velocities relevant for characterizing the orbital dynamics
and geometry of the WD+WD progenitor binary.

\section{The Donor Radius}
\label{s.radius}

The D$^6$ scenario predicts that the donor star is a Roche-filling WD
when its companion explodes as a supernova and ejects it from the
system as a hypervelocity runaway. This places a strong constraint on
the donor WD radius as a function of its measured velocity.
We define the mass ratio of the system as
$q \equiv M_{\rm don}/M_{\rm acc}$, where $M_{\rm don}$ is the mass of
the donor WD that will become the observed D$^6$ runaway, and
$M_{\rm acc}$ is the mass of the accretor WD that explodes as a
Type~Ia SN.
The orbital velocity $v_{\rm orb}$ of the donor star about the center
of mass of the binary system can then be expressed as
\begin{equation}
v_{\rm orb}^2 = \frac{G M_{\rm acc}}{a(1+q)}~,
\end{equation}
where $a$ is the orbital separation.
The Roche lobe radius $R_{RL}$ of the donor WD can be expressed using the
\cite{Eggleton1983} approximation:
\begin{equation}
\label{eq:eggleton}
R_{RL} = \frac{0.49 q^{2/3}a}{0.6 q^{2/3} + \ln(1+ q^{1/3})}~.
\end{equation}
The fact that the WD donor fills
its Roche lobe ($R_{\rm don} = R_{RL}$) at the time of SN detonation
allows us to combine the above equations to express the donor radius as
\begin{equation}
\label{eq:Rdonor}
R_{\rm don} = \frac{0.49 q^{2/3}G M_{\rm acc}}{v_{\rm orb}^2(1+q)[0.6 q^{2/3} + \ln(1+ q^{1/3})]}~.
\end{equation}
This expression for the donor radius depends only on the orbital
velocity of the donor and the masses of the two stars in the
binary. In the D$^6$ scenario for producing a Type~Ia SN, the accretor
is expected to be a C/O WD with mass in the range
$M_{\rm acc} \in (0.85\,M_\odot, 1.15\,M_\odot)$.
We select this accretor mass range to cover the full range of plausible
D$^6$ scenarios for producing normal Type~Ia SNe
\citep{Blondin2017,Shen2021a,Shen2021}.

\section{Mass Constraints}
\label{s.mass}

\begin{figure}
  \centering
  \includegraphics[width=\apjcolwidth]{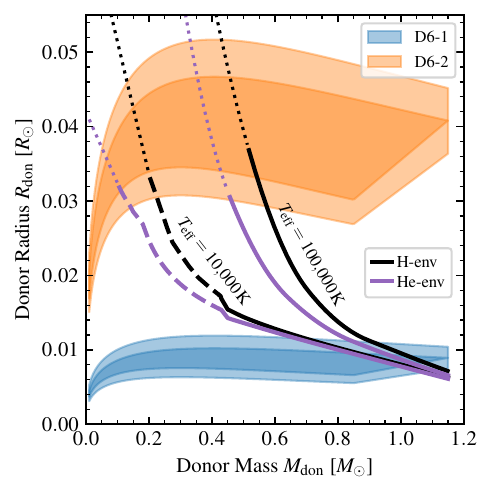}
  \caption{Donor radii according to
    Equation~\eqref{eq:Rdonor} compared to WD mass-radius relations.
    The solid portions of the curves represent the mass-radius
    relation given by the \cite{Bedard2020} C/O
    models. The sections of the mass-radius relations covered by models
    with He cores are indicated by the dashed portions of the
    curves, and regions where we are extrapolating are
    indicated by the dotted portions. Models with H envelopes are
    represented by black curves (H-env), and models with He envelopes
    are represented by purple curves (He-env).
    For each D$^6$ object, the inner shaded band represents possible radius
    solutions using the median posterior velocity with a range of possible
    accretor masses assuming $M_{\rm don} < M_{\rm acc}$ and
    $M_{\rm acc} \in (0.85\,M_\odot, 1.15\,M_\odot)$.
    The outer shaded bands represent the additional radius uncertainty
    introduced by the $\approx 1\sigma$ velocity uncertainty.
    D6-3 looks very similar to D6-1 on this plot except that it has a
    much larger velocity uncertainty, so we only display D6-1 and D6-2
    here for clarity.
  }
  \label{fig:MR}
\end{figure}

The fact that the donor star in the D$^6$ scenario must be a WD
implies that the donor radius must also fall along the
WD mass-radius relation, which is independent of the radius estimate
provided in Equation~\eqref{eq:Rdonor}. The measured velocities of the
D$^6$ objects therefore provide a constraint on their possible masses
as donor stars as shown in Figure~\ref{fig:MR}, where we plot the
range of possible radii according to Equation~\eqref{eq:Rdonor} with
$M_{\rm acc} \in (0.85\,M_\odot, 1.15\,M_\odot)$ and $v_{\rm orb}$
from the {\it Gaia} EDR3 measurements described in
Section~\ref{s.velocities}, along with WD mass-radius relations
representing the full range of potential WD temperatures and 
envelope configurations.

\begin{figure*}
  \centering
  \includegraphics[width=\apjcolwidth]{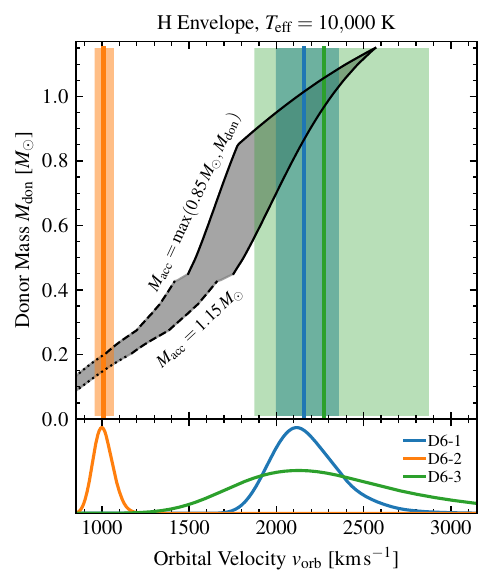}
  \includegraphics[width=\apjcolwidth]{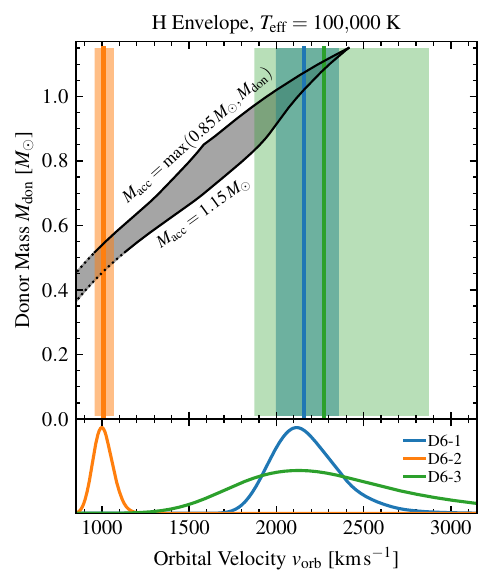}
  \caption{Solutions for WD donor mass as a function of velocity
    assuming a cool WD ($T_{\rm eff} = 10{,}000\,\rm K$) mass-radius
    relation in the left panel and a hot WD
    ($T_{\rm eff} = 100{,}000\,\rm K$) relation in the right
    panel. These mass-radius relations include H envelopes in both cases (see Figure~\ref{fig:MdonorHe} for the He envelope case).
    The gray band gives the possible range of solutions for
    accretor masses in the range
    $M_{\rm acc} \in (0.85\,M_\odot, 1.15\,M_\odot)$, subject to the
    constraint that $M_{\rm don} < M_{\rm acc}$. The blue, orange, and
    green shaded regions give the $\pm 1\sigma$ velocity ranges for D6-1,
    D6-2, and D6-3 respectively, and the corresponding vertical lines
    mark the median posterior velocity for each object.
    The lower panels show the {\it Gaia} EDR3 velocity posterior
    distributions from Figure~\ref{fig:velocities} (arbitrary scale).}
  \label{fig:Mdonor}
\end{figure*}

\begin{figure*}
  \centering
  \includegraphics[width=\apjcolwidth]{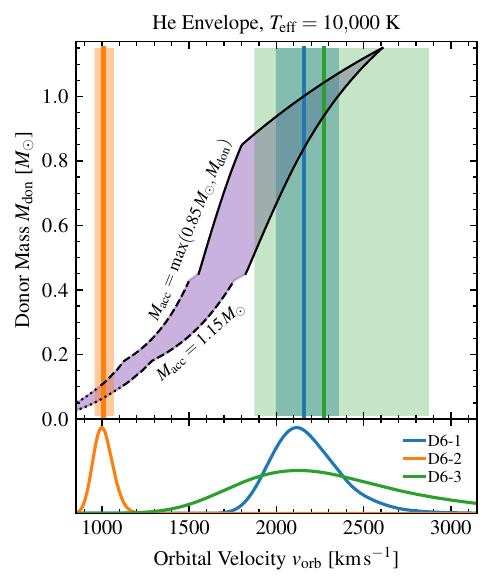}
  \includegraphics[width=\apjcolwidth]{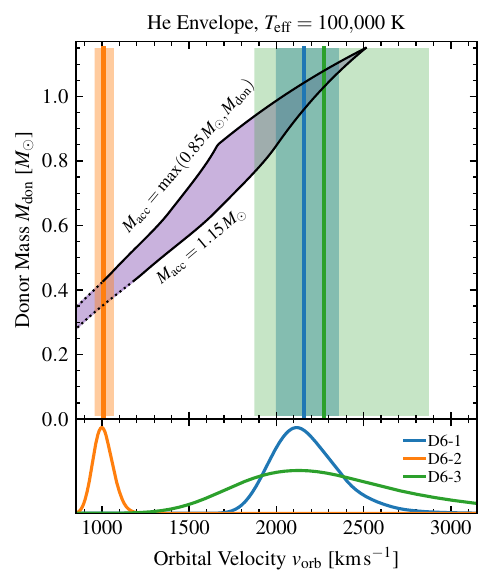}
  \caption{Same as Figure~\ref{fig:Mdonor} but using the mass-radius relations for WDs with no H envelope, i.e.\ with He envelopes.} 
  \label{fig:MdonorHe}
\end{figure*}

\subsection{WD Mass-Radius Relations}

The donor WD in the D$^6$ scenario may have a thin outer H envelope when
it first fills its Roche lobe and begins to transfer mass. Depending on
the mass ratio and compactness of this envelope, dynamically unstable mass
transfer may or may not begin right away. If mass transfer immediately proceeds
to a dynamically unstable runaway, then the WD will respond to mass
loss adiabatically and expand to fill its Roche lobe even as the H envelope
is quickly stripped away, so that the radius including this envelope
determines the orbital velocity at the time of
instability and SN detonation. On the other hand, the H envelope may
transfer stably for a time.
If the envelope can maintain thermal equilibrium, the orbit may continue to shrink
toward the more compact radius of the underlying He envelope before
dynamical instability sets in. We therefore present mass-radius relations
for WD models both with and without H envelopes, which represent an upper
and lower bound on the relevant radius of a WD at a given mass and temperature.
In reality, some WD+WD systems may fall somewhere between these two limiting
cases once mass transfer sets in before ultimately experiencing
instability \citep{Shen2015}. Figure~\ref{fig:MR} shows that the presence of an H envelope
has only a small impact on the overall radius of massive WDs, so it will not
have a large impact on mass inferences in that regime, but for lower mass
WDs it can substantially influence the result.

WDs in compact binaries may also experience tidal heating that injects energy into the WD interior or envelope and causes it to be inflated relative to the radius of a cooler WD of the same mass \citep{Fuller2012,Piro2019}. We use $T_{\rm eff}$ as a proxy for this inflation in Figure~\ref{fig:MR}, since hotter WD models also have more inflated radii, and $T_{\rm eff}$ is easily accessible in the grids of WD models that we use to construct mass-radius relations \citep{Althaus2013,Bedard2020}. In reality, the interior structure of a tidally heated WD may be more complicated than the structure of a WD cooling model that might produce a similar radius and effective temperature, but $T_{\rm eff}$ provides a loose representation of the possible amount of inflation that an extremely tidally heated WD might achieve.

For models that include H envelopes, we construct mass-radius relations for
both cool ($T_{\rm eff} = 10{,}000\,\rm K$) and hot ($100{,}000\,\rm K$) donor WDs
using the models of \cite{Althaus2013} and \cite{Bedard2020} as follows:
\begin{itemize}
\item For $M_{\rm don} \geq 0.45\,M_\odot$, we interpolate from the C/O WD
  models of \cite{Bedard2020} using Sihao Cheng's {\tt WD\_models} package.%
  \footnote{\url{https://github.com/SihaoCheng/WD_models/releases/tag/v1.1}}
  We adopt the thick hydrogen envelope models in this regime, so these
  represent an upper limit for the radius at a given mass, which will
  translate into a slightly slower velocity at Roche lobe overflow.
  It is also possible that donor WDs of mass $M_{\rm don}
  \gtrsim 1.05\,M_\odot$ would have O/Ne cores that are slightly more
  compact than the mass-radius relation shown on the high mass end
  here. However, the more massive accreting WD in the D$^6$ scenario
  should have a C/O core to explode as a normal Type~Ia SN, and the
  donor WD needs to be less compact than the accretor, so we adopt the
  C/O mass-radius relation even for the most massive possible
  donors. In any case, O/Ne models would only be slightly more compact
  and therefore would not achieve qualitatively higher orbital
  velocities before mass transfer.
\item For $M_{\rm don} < 0.45\,M_\odot$, we interpolate based on the
  helium-core WD model grid of \cite{Althaus2013}, which reliably
  covers a mass range of about $0.2$--$0.43\,M_\odot$ for
  $T_{\rm eff} \leq 30{,}000\,\rm K$.
  It is also possible that a significant fraction of WDs
  in compact binaries with
  $M_{\rm don} \approx 0.3$--$0.5\,M_\odot$ could descend from hot
  subdwarf stars and would therefore have C/O cores rather than He cores
  \citep{Zenati2019,Kupfer2020a,Kupfer2020b,Bauer2021,Schwab2021,Romero2021}.
  For simplicity, we do not include this possibility in our plots in
  this paper, but it should be noted that some WDs in this mass
  range could have somewhat more compact cores than the He models
  presented here. He cores below $0.3\,M_\odot$ cannot sustain He
  burning, so the lowest mass WDs matching the radii that we infer for
  D6-2 in Figure~\ref{fig:MR} should reliably have He cores.
\item For the lower-mass regions where we do not have model coverage,
  we extrapolate from the lowest mass for which we have reliable
  models, as shown by the dotted curves in Figure~\ref{fig:MR}. For
  $T_{\rm eff} = 100{,}000\,\rm K$, there is no coverage from the
  lower-mass \cite{Althaus2013} model grid because He-core WDs
  generally do not reach this high temperature, but the C/O
  models of \cite{Bedard2020} have radii that extend up to roughly the
  region of interest for $M_{\rm don} \approx 0.5\,M_\odot$, and a
  mild extrapolation of the C/O models to lower masses covers the
  entire relevant radius region. For the lower-temperature models, we
  extrapolate the He-core model relation when $M_{\rm don} \lesssim
  0.2\,M_\odot$.
\end{itemize}

We follow a similar approach to construct the mass radius relation
for models with no H envelope,
using the DB WD models of \cite{Bedard2020} for C/O WDs when
$M_{\rm don} \geq 0.45\,M_\odot$. For the lower-mass regime where we
expect WDs to have He cores, we construct our own model
grid using {\tt MESA} version 15140
\citep{Paxton2011,Paxton2013,Paxton2015,Paxton2018,Paxton2019}.%
\footnote{Our {\tt MESA} models are available at\\
    \url{https://doi.org/10.5281/zenodo.5721766}.}
We construct these models starting with a $1.1\,M_\odot$ star
that ascends the RGB until its He core reaches a mass of $0.35\,M_\odot$.
We then remove all of the outer hydrogen envelope to form a semi-degenerate
He ball of nearly uniform composition, representing a proto-He WD with
no H envelope. To make a grid of He WD models, we use a relaxation process
to rescale this model to masses in the range $0.1$--$0.43\,M_\odot$, and
then let the models cool to $10{,}000\,\rm K$ to form a grid of He WDs
suitable for a mass-radius relation.

Once again, for the hotter ($100{,}000\,\rm K$) set of models, He WDs
do not generally reach this temperature during isolated evolution, so we use the \cite{Bedard2020}
DB WD models for $M_{\rm don} \geq 0.45\,M_\odot$, and we show a rough extrapolation
of that mass-radius relation for lower WD masses at this temperature.

\subsection{Donor Mass from Measured Velocity}

The donor mass can be expressed as a function of donor velocity by
setting $R_{\rm don}$ from Equation~\eqref{eq:Rdonor} equal to the WD
mass-radius relation. We show results based on WD mass-radius relations at
both cool and hot effective temperatures in Figure~\ref{fig:Mdonor} for models that include H envelopes, and in Figure~\ref{fig:MdonorHe} for models that have no H envelope.
It is immediately clear that D6-1 and D6-3 require rather massive
donor WDs to produce their high velocities, with their median
velocities favoring $M_{\rm don} \approx 1.0\, M_\odot$
depending on $T_{\rm eff}$ and envelope composition.

WDs in binaries should generally have ample time to cool to
$T_{\rm eff} \approx 10{,}000\,\rm K$ before coming into contact due to
gravitational wave radiation and beginning mass transfer (e.g.\ \citealt{Maoz2018,Cheng2020}). The left
panels of Figures~\ref{fig:Mdonor} and~\ref{fig:MdonorHe} would then suggest that D6-2 was an
extremely low-mass (ELM) WD with a He core and
$M_{\rm don} \lesssim 0.2\, M_\odot$ to explain its low observed
velocity. On the other hand, the right panels show that D6-2 is also marginally consistent with
having been a very hot $\approx 0.5\, M_\odot$ C/O WD that was
inflated when mass transfer began due to its high temperature. This
configuration could possibly be produced by tidal heating in the
compact WD binary as it approached contact
\citep{Fuller2012,Burdge2019,Piro2019}, though theoretical predictions for the
total amount of tidal heating and corresponding inflation in the outer
layers are complex and still uncertain in this regime
\citep{Burkart2013,Burkart2014,Fuller2013,Fuller2014,Yu2020}.
According to \cite{Piro2019}, the total luminosity available from tidal
heating can be on the order of $\approx 1\,L_\odot$, which could translate
into a temperature of $T_{\rm eff} \approx 30$--$40{,}000\,\rm K$ for a WD
with the radius we calculate for D6-2. This could contribute somewhat
toward the inflation of a WD in a binary system like D6-2. However,
in the $P_{\rm orb} = 6.9$~minute WD+WD system ZTF~J1539+5027 observed by
\cite{Burdge2019}, the lower mass WD that will eventually become the donor
is very cool and inconsistent with significant tidal heating, while the
higher mass WD in that system is much hotter.
\cite{Burdge2019} also argue that it is difficult for tidal heating to
explain a temperature much hotter than $40{,}000\,\rm K$ for a WD+WD system
at this orbital period.

Tidal heating has a much smaller impact on mass inferences for D6-1 and D6-3, where it only pushes to slightly higher masses above $\approx 1.0\, M_\odot$ according to the right panels in Figures~\ref{fig:Mdonor} and~\ref{fig:MdonorHe}.

\section{Rotation}
\label{s.rotation}

\cite{Chandra2021} have recently identified a period for the D6-2
remnant that most likely corresponds to its current rotation period.
The current radius of the D6-2 donor remnant is significantly inflated
compared to its radius prior to the SN companion explosion, and
\cite{Chandra2021} discuss the relationship between the current observed
rotation rate and the potential tidally locked rotation rate of the
D6-2 donor in the binary just prior to SN detonation.
The required radius for D6-2 in Figure~\ref{fig:MR}
further refines the possible inferences that can be
made about rotation in the D6-2 system. Assuming tidal locking
for D6-2 as a donor star, conserving angular momentum in its outer
layers predicts that its current rotation period should be
\begin{equation}
  P_{\rm rot} \approx P_{\rm orb} \left( \frac{R_{\rm obs}}{R_{\rm don}} \right)^2~,
  \label{eq:Prot}
\end{equation}
where $R_{\rm obs} = 0.2\,R_\odot$ is the current measured radius of
the D6-2 remnant object according to \cite{Chandra2021}. The system
likely would have had an orbital period of
$P_{\rm orb} = 3$--$7$~minutes at the moment of SN detonation, and
therefore Equation~\eqref{eq:Prot} and Figure~\ref{fig:MR} predict
that the current observed rotation period should be
$P_{\rm rot}\approx 2$--$5$~hours. However,
\cite{Chandra2021} report a measured rotation period of
$P_{\rm rot}\approx 15.4$~hours. This may require that D6-2
experienced some angular momentum loss after interacting with SN
ejecta to reach its current rotation period. Alternatively, because
D6-2 was so much less compact than the other D$^6$ objects, its more
loosely bound surface layers could have experienced significant
stripping when interacting with SN ejecta. The current surface could
have originally been deeper material at a smaller radius that
inflated more to reach the current observed radius,
so that Equation~\eqref{eq:Prot} would predict a longer rotation
period than the value given by using $R_{\rm don}$.

\section{Discussion}
\label{s.discussion}

We have calculated WD donor masses for D$^6$ runaways using observed {\it Gaia} EDR3 velocities and WD mass-radius relations for several combinations of possible envelope configurations and effective temperatures.
Our conclusions on the high donor mass ($M_{\rm don} \gtrsim 0.8\, M_\odot$) for both D6-1 and D6-3 hold across all effective temperatures and non-degenerate envelope layers we sample. Since the donor must be less massive than the accretor, this implies that two of the three known D$^6$ stars had progenitor binaries composed of two massive white dwarfs, both likely $>0.8\, M_\odot$. No massive WDs are currently known to show large radial-velocity changes indicating another massive companion, though such searches are challenging due to their intrinsic faintness \citep{Rebassa-Mansergas2019}. Additionally, most radial-velocity surveys have limited numbers of massive white dwarfs; just 31 out of 643 targets ($<5\%$) in the ESO SN Ia Progenitor surveY (SPY) have a mass $>0.8\, M_\odot$ \citep{Napiwotzki2020}.
If WD+WD systems containing two massive WDs are indeed intrinsically rare, this may suggest that the D$^6$ channel only contributes a fraction of explosions to the overall thermonuclear supernova rate.

The low donor mass of the D6-2 system in Figure~\ref{fig:Mdonor}
implies that it would have originated in a system with a low mass
ratio. A system with such a low mass ratio would not necessarily
experience dynamical instability at the onset of mass transfer
\citep{Marsh2004,Dan2012}. However, this dynamical instability is a
prerequisite for producing the detonation that would eventually lead
to the liberation of the donor as a hypervelocity runaway. Therefore,
the existence of the runaway D6-2 donor remnant
lends further support to the arguments of
\cite{Shen2015} and \cite{Brown2016} that most low mass ratio
double WD binaries are disrupted at contact rather than experiencing
stable mass transfer from an ELM WD to form an AM~CVn system (e.g.\ \citealt{Wong2021}).

Alternatively, the qualitatively slower velocity of D6-2 compared to the other D$^6$
candidates may instead suggest that it had a different compact binary
evolution origin. For example, the highest velocity tail of the
runaway velocity distribution for He star donors to WD companions
calculated by \cite{Neunteufel2020} and \cite{Neunteufel2021} could be
marginally compatible with the velocity of D6-2.
This could require an unlikely random alignment of orbital ejection
with a halo star orbit, as has been suggested for the runaway He star
US~708 \citep{Geier2015,Brown2015,Bauer2019,Liu2021}.
However, \cite{Hermes2021} have argued that relatively slow rotation ($P_{\rm rot} \gtrsim 10$~hours) is problematic for the runaway He star donor channel, and it is also somewhat difficult to explain for a runaway WD donor as discussed in Section~\ref{s.rotation}.
Instead, D6-2 could plausibly be related to the LP~40--365 class of
runaway stars, which have been suggested to be bound remnants of
Type~Iax SNe \citep{Jordan2012,Vennes2017,Raddi2018mnras,Raddi2018apj,Raddi2019,Hermes2021}.
The three candidates identified as part of that class so far have
velocities somewhat below $1000\,\rm km\,s^{-1}$, but the
distributions of kick velocities in that scenario are uncertain, and
predictions for final velocities of remnants are not constrained well
enough to completely exclude a velocity as high as that of D6-2.
\cite{Pakmor2021} have also proposed yet another possibility
    for producing runaways with velocities up to $1000$--$1500\,\rm km\,s^{-1}$
    from binaries containing a massive He-C/O WD
    (e.g. \citealt{Zenati2019,Pelisoli2021}) donating material onto a C/O
    WD companion. In that case, the donor's thick He envelope may experience
    a detonation and release the {\it accretor} WD as a hypervelocity
    runaway. This speculative scenario may achieve velocities 
    sufficient to explain D6-2, but not D6-1 or D6-3.

In conclusion, our calculations in this work have enabled new inferences
about the masses of the observed D$^6$ runaways that fall on either end
of the WD mass spectrum, while objects near the 0.6$\,M_\odot$ peak of
the WD mass distribution are absent so far.
The relatively low velocity of D6-2 points to a low donor WD mass, introducing new
puzzles for its interpretation as a D$^6$ donor remnant in the context
of growing observational data about its present-day properties,
such as surface rotation and circumstellar material \citep{Chandra2021}.
On the other end of the velocity scale,
D6-1 and D6-3 both consistently point toward origins in massive
WD+WD binaries regardless of modeling assumptions.
This clearly motivates further observational followup work,
especially detailed spectroscopic analysis to understand the complicated
surface compositions and structures of D$^6$ stars.

\begin{acknowledgments}
    We thank Warren Brown and Charlie Conroy for comments and suggestions.
    VC is supported in part by the James Mills Peirce fellowship at Harvard University. Financial support for KJS and JJH was in part provided by NASA/ESA {\em Hubble Space Telescope} program \#15871. KJS also received support from NASA through the Astrophysics Theory Program (NNX17AG28G). The computations in this paper were run on the FASRC Cannon cluster supported by the FAS Division of Science Research Computing Group at Harvard University.
\end{acknowledgments}

\bibliographystyle{yahapj}

\end{document}